\UseRawInputEncoding
\documentclass{article}
\usepackage{arxiv}

\usepackage{graphicx,amsfonts,amssymb,amsmath,mathrsfs,hyperref,bm,dsfont}
\usepackage[normalem]{ulem}
\usepackage{amsfonts} 
\usepackage{lscape}
\usepackage{amsmath}
\usepackage{mathtools}
\usepackage{xcolor}     
\usepackage{array} 
\newcolumntype{H}{>{\iffalse}c<{\fi}@{}}
\usepackage{lipsum}
\usepackage{amssymb}
\usepackage{multicol}
\usepackage{bm}
\usepackage{amsthm}
\usepackage{algorithm}
\usepackage{algpseudocode}
\usepackage{import}
\usepackage{tikz}
\usetikzlibrary{arrows.meta}
\usepackage{xifthen}
\usepackage{caption} 
\newcommand{\E}{\mathbb{E}}
\newcommand{\R}{\mathbb{R}}
\newcommand{\Q}{\mathbb{Q}}

\newcommand{\ud}{\,\mathrm{d}}

\usepackage{authblk}

\newtheorem{theorem}{Theorem}

\newtheorem{proposition}[theorem]{Proposition}

\usepackage{color}

\usepackage{setspace}

\usepackage{hyperref}
\colorlet{myred}{red!80!black}
\colorlet{myblue}{blue!80!black}
\colorlet{mygreen}{green!60!black}
\hypersetup{
	colorlinks=true,
	linkcolor={myblue!70},
	citecolor={mygreen!70},
	urlcolor={myred!70}
}
\def\be{\begin{equation}}
\def\ee{\end{equation}}
\def\bea{\begin{eqnarray}}
\def\eea{\end{eqnarray}}

\begin{document}

\title{Application of Tensor Neural Networks to Pricing Bermudan Swaptions}

\author[1,5]{\thanks{Email: raj.patel@multiversecomputing.com}\hspace{0.5mm} Raj G. Patel}
\author[1,5]{Tomas Dominguez}
\author[1]{Mohammad Dib}

\author[1]{Samuel Palmer}

\author[3]{Andrea Cadarso}
\author[3]{Fernando De Lope Contreras}
\author[3]{Abdelkader Ratnani}
\author[3]{Francisco Gomez Casanova}

\author[4]{Senaida Hernández-Santana}
\author[4]{Álvaro Díaz-Fernández}
\author[4]{Eva Andrés}
\author[4]{Jorge Luis-Hita}
\author[4]{Escolástico Sánchez-Martínez}

\author[1]{Samuel Mugel}
\author[2]{Román Orús}

\affil[1]{%
Multiverse Computing, Centre for Social Innovation, 192 Spadina Ave, Suite 509, Toronto, M5T 2C2, Canada} 
\affil[2]{%
Multiverse Computing, Paseo de Miram\'on 170, 20014 San Sebasti\'an, Spain} 
\affil[3]{%
BBVA Corporate \& Investment Banking, Calle Sauceda 28, 28050 Madrid, Spain} 
\affil[4]{%
BBVA Quantum, Calle Azul 4, 28050 Madrid, Spain}
\affil[5]{%
University of Toronto, Toronto, Ontario M5S 2E4, Canada} 

\maketitle
\begin{abstract}

The Cheyette model is a quasi-Gaussian volatility interest rate model widely used to price interest rate derivatives such as European and Bermudan Swaptions for which Monte Carlo simulation has become the industry standard. In low dimensions, these approaches provide accurate and robust prices for European Swaptions but, even in this computationally simple setting, they are known to underestimate the value of Bermudan Swaptions when using the state variables as regressors. This is mainly due to the use of a finite number of pre-determined basis functions in the regression. Moreover, in high-dimensional settings, these approaches succumb to the Curse of Dimensionality. To address these issues, Deep-learning techniques have been used to solve the backward Stochastic Differential Equation associated with the value process for European and Bermudan Swaptions; however, these methods are constrained by training time and memory. To overcome these limitations, we propose leveraging Tensor Neural Networks as they can provide significant parameter savings while attaining the same accuracy as classical Dense Neural Networks. In this paper we rigorously benchmark the performance of Tensor Neural Networks and Dense Neural Networks for pricing European and Bermudan Swaptions, and we show that Tensor Neural Networks can be trained faster than Dense Neural Networks and provide more accurate and robust prices than their Dense counterparts.

\end{abstract}

\section{Introduction}

Partial Differential Equations (PDE) are a crucial tool for modeling a variety of problems in Quantitative Finance. Many of these problems, such as the option pricing problem, can be recast as solving a parabolic PDE. Classical methods for solving such PDE involve mesh-based or Monte-Carlo approaches. Unfortunately, these techniques fail to scale to high-dimensional settings due to their dependence on spatio-temporal grids and an exorbitant number of sample paths. Recent advances in the field of Deep Learning \cite{lecun2015deep} have made it possible to address these issues by approximating the unknown solution to a PDE using a Dense Neural Network (DNN) \cite{huang2022partial, Raissi, Beck_2019, Han_2018, E_2017, patel2022quantuminspired}. This breakthrough has had deep implications for the field of Quantitative Finance, where it has made it possible to consider models with a previously computationally intractable number of assets or agents without making any provisional assumptions on their correlation or interaction structure. The flexibility and omnipresence of PDE makes these advances equally relevant for the domains of high-dimensional Stochastic Optimal Control and Forward-Backward Stochastic Differential Equations, both of which can be recast in the language of PDE \cite{Cheridito}. In particular, these Deep Learning techniques can be leveraged to address problems such as option pricing, optimal liquidation, portfolio optimization and systemic risk in quantitative finance \cite{jaimungal, ABOUSSALAH2020112891}, but they can also be extended to industries as diverse as energy management or supply chain and inventory control.

This paper explores the problem of pricing European and Bermudan Swaptions \cite{hull, wilmott, mcdonald2013derivatives} in the Cheyette model \cite{cheyette2001markov}, a Markovian quasi-Gaussian volatility interest model. The importance of this problem stems from the considerable trading volume of interest rate derivatives, including European and Bermudan Swaptions, and the wide-spread use of the Cheyette model in industry. Classically, these interest rate derivatives are priced by Monte-Carlo methods. In the case of Bermudan Swaptions, these Monte-Carlo approaches rely on the regression-based Longstaff-Schwartz method \cite{l-s-algo} which becomes computationally intractable in high dimensions when taking the factors as the state variables in the regression. In addition to suffering from the Curse of Dimensionality, this regression-based method is widely known to under-estimate the true price of the option \cite{Glas04}. There have been attempts industrially to leverage Principal Component Analysis to first extract relevant factors, and then solve their associated PDE; however, with five or more factors, the PDE cannot be solved by traditional approaches such as finite difference or finite element methods. To circumvent these problems, in this paper, the problem of pricing European and Bermudan Swaptions is recast as the problem of solving a system of forward-backward Stochastic Differential Equations (SDE) \cite{Cheridito}. Efficient methods for approximating the solutions to forward-backward SDE using Dense Neural Networks have recently been proposed \cite{Raissi-part1,Raissi-part2, Patel_2022}; however, in spite of their apparent success, Dense Neural Network approaches are computationally expensive and limited by memory \cite{TNN_NIPS, tn_memory, xue13_interspeech}. Furthermore, there have been recent attempts to address the problem of solving PDE using the advances in Quantum Computing \cite{garciaripoll, li2023quantum, jin2022quantum}. While promising, given the current limitations of quantum algorithms such as the limited number of qubits and high error rates, these approaches require important hardware advances before becoming a viable tool to solve high-dimensional PDE. In this paper we take a quantum-inspired approach by combining Tensor Networks with Dense Neural Networks to obtain Tensor Neural Networks (TNN).

Originally developed in the field of physics to describe strongly-correlated structures, Tensor Networks \cite{RomanTN, sukhi_1, tnn_def} have gained popularity in Machine Learning due to their efficiency in describing high-dimensional vectors and operators. Tensor Networks have been successfully applied to various Machine Learning tasks \cite{CACIB_1, TNN_Survey} such as classification \cite{NIPS2016_6211,Stoudenmire_2018,glasser2018supervised,efthymiou2019tensornetwork,bhatia2019matrix,Liu_2019,9058650}, generative modeling \cite{PhysRevX.8.031012,PhysRevB.99.155131,PhysRevB.101.075135}, sequence modeling \cite{bradley2020modeling}, quantum chemistry \cite{quantum_chemistry}, dimensionality reduction \cite{dimensionality} and subspace learning \cite{subspace} to mention just a few. Inspired by recent works, we propose this transformation from a Dense Neural Network into a Tensor Neural Network for the case of Swaptions, resulting in improved training performance as compared to the classical approaches and reduced memory consumption. To back this claim, a detailed comparison of Tensor Neural Networks against the best performing Dense Neural Network with the same number of parameters is performed. This is supported by contrasting a given Tensor Neural Network with the Dense Neural Network with the same number of neurons, and therefore considerably more parameters \cite{CACIB_1, patel2022quantuminspired}.

It should be noted that the strategy proposed to price European Swaptions differs from that put forth to price Bermudan Swaptions. In the European setting, a forward methodology leveraging a single Neural Network to solve the SDE associated with the Swaption value is used. On the other hand, in the Bermudan setting, a stacked sequence of Neural Networks, one for each exercise time, is exploited. This requires a careful handling of the early-exercise nature of the option which is done by starting at the terminal time and recursively propagating the terminal condition backwards in time while simultaneously learning the optimal decision boundary. Despite these deep and fundamental strategic differences, both of these problems benefit from the computational speed-up associated with Tensor Neural Networks.

\noindent
This paper is organized as follows. In Section \ref{sec:Problem_Formulation}, the concept of European and Bermudan Swaptions is briefly reviewed. Following this, the Cheyette Model is defined, and then the pricing of European and Bermudan Swaptions in the Cheyette Model is discussed. Section \ref{sec:NN_for_swaptions} focuses on the Neural Network setup required to price European and Bermudan Swaptions in the Cheyette model. The notion of tensorizing a Neural Network is then briefly reviewed in Section \ref{sec:TNN}. The results of numerical experiments using the methods introduced in this paper to price European and Bermudan Swaptions are presented and discussed in Section \ref{sec:results}, where it is shown that Tensor Neural Networks outperform Dense Neural Networks. Finally, in Section \ref{sec:conclusion}, conclusions and further areas of investigation are put forth.

\newpage
\section{Problem Formulation and Mathematical Framework}\label{sec:Problem_Formulation}

In this section we define the two financial instruments, the \emph{European Swaption} and the \emph{Bermudan Swaption}, whose no-arbitrage price we determine in this paper. We also introduce the stochastic interest rate model that will drive the evolution of these financial derivatives, and discuss how to price these instruments under said model.

\subsection{European and Bermudan Swaptions}

A \emph{European Swaption} is a financial contract that gives the holder the right, but not the obligation, to enter the \emph{payer leg} of an \emph{interest rate swap} with a pre-determined \emph{tenor structure} $0<T_0 < T_1 < \ldots < T_n$ and fixed rate $K\geq 0$ at the future time $T_0$. A \emph{Bermudan Swaption} is a financial contract that gives the holder the right, but not the obligation, to enter the \emph{payer leg} of an \emph{interest rate swap} with a pre-determined \emph{tenor structure} $0<T_0 < T_1 < \ldots < T_n$ and fixed rate $K\geq 0$ at any future tenor date $T_m$ with $0\leq m\leq n$ An analogous discussion to what follows can be developed in the case that the holder of the Swaption enters the \emph{receiver leg} of the interest rate swap, but, for brevity, we focus exclusively on payer Swaptions. An \emph{interest rate swap} with tenor structure $0<T_0<T_1<\ldots<T_n$ and fixed rate $K\geq 0$ is a financial contract in which one party, the \emph{payer}, pays another party, the \emph{receiver}, the fixed cash-flow $K$ in exchange of a floating cash flow $\ell_m$ at each future time $T_m$ in the tenor structure. The floating rate $\ell_m$ between times $T_{m-1}$ and $T_m$ can be expressed in terms of the over-night risk-free-rate at $T_m$ and the value $P(T_{m-1},T_m)$ of a \emph{zero-coupon bond} with maturity $T_m$ at time $T_{m-1}$. Recall that a \emph{zero-coupon bond} with maturity $T>0$ is a contract which guarantees the holder one dollar to be paid at time $T>0$, and its price is the stochastic process denoted by $(P(t,T))_{t\in [0,T]}$. Zero-coupon bonds are the building blocks of any interest rate theory, and their stochastic evolution will be the driver for the value of the European and Bermudan Swaptions that we price. 

\subsection{The Cheyette model}
\label{sec:cheyette}
Throughout this paper we will place ourselves in the context of the \emph{Cheyette model} \cite{cheyette2001markov}. The Cheyette model is a special case of the \emph{Heath-Jarrow-Morton} (HJM) stochastic interest rate model, so, instead of directly modeling the evolution of zero-coupon bonds, we model the evolution of the forward curve
\begin{equation}\label{e.f.from.P}
f(t,T):=-\frac{\partial}{\partial T}\log P(t,T).
\end{equation}
More specifically, we assume that the \emph{initial forward curve} $T\mapsto f(0,T)$ is known and evolves according to the real-world dynamics
\begin{equation}
\ud f(t,T)=\mu(t,T)\ud t+\sigma(t,T)\cdot \ud W(t)
\end{equation}
for some drift $\mu:\R_{\geq 0}\times \R_{\geq 0}\to \R$, some volatility $\sigma: \R_{\geq 0}\times \R_{\geq 0}\to \R^d$ and some $d$-dimensional Brownian motion $W$. Notice that the evolution of zero-coupon bonds may be recovered from the evolution of the forward curve,
\begin{equation}\label{e.p.to.f}
P(t,T)=\exp\bigg(-\int_t^T f(t,s)\ud s\bigg).
\end{equation}
Combining Ito's lemma with the Girsanov theorem, under appropriate regularity assumptions, it is possible to find a unique risk-neutral measure $\Q$ under which all discounted zero-coupon bonds are martingales. Here the discounting is done relative to the \emph{short rate}
\begin{equation}
r(t):=f(t,t).
\end{equation}
Under this risk-neutral measure, the initial forward curve evolves according to the stochastic dynamics
\begin{equation}\label{e.HJM.f.under.Q}
\ud f(t,T)=\sigma(t,T)\cdot \sigma^*(t,T)\ud t +\sigma(t,T)\cdot \ud W^\Q(t)
\end{equation}
for some $d$-dimensional $\Q$-Brownian motion $W^\Q$ and the functions
\begin{equation}\label{e.sigma*}
\mu^*(t,T):=\int_t^T \mu(t,s)\ud s \quad \text{and} \quad \sigma^*(t,T):=\int_t^T \sigma(t,s)\ud s.
\end{equation}
Since we are concerned with the \emph{pricing} of financial derivatives, by the HJM model we will mean the stochastic evolution \eqref{e.HJM.f.under.Q} of the forward curve under the risk-neutral measure $\Q$.

One of the main difficulties in analyzing the HJM model is the path dependence of the process $\sigma^*$ in \eqref{e.sigma*} which makes the forward curve non-Markovian. A classical way to overcome this difficulty is to impose that the volatility process $\sigma=(\sigma_i)_{i\leq d}$ be separable,
\begin{equation}
\sigma_i(t,T):=h_i(t,X_i(t))g_i(T)
\end{equation} for some deterministic functions $h=(h_i)_{i\leq d}$ and $g=(g_i)_{i\leq d}$ and some adapted process $X=(X_i)_{i\leq d}$. A direct calculation reveals that under this assumption the forward curve is Markovian,
\begin{equation}\label{e.cheyette.f}
f(t,T)=f(0,T)+\sum_{i\leq d}\frac{g_i(T)}{g_i(t)}\bigg(X_i(t)+Y_i(t)\int_t^T \frac{g_i(v)}{g_i(t)}\ud v\bigg)
\end{equation}
for the stochastic processes $X=(X_i)_{i\leq d}$ and $Y=(Y_i)_{i\leq d}$ defined by
\begin{align}
X_i(t)&:=g_i(t)\int_{0}^t h^2_i(s,X_i(s))\int_s^t g_i(v)\ud v\ud s+g_i(t)\int_{0}^t h_i(s,X_i(s))\ud W_i^\Q(s),\\
Y_i(t)&:=g_i(t)^2\int_{0}^t h_i^2(s,X_i(s))\ud s.
\end{align}
Writing $\odot$ for the Hadamard product on the space of matrices and applying Ito's lemma shows that the factor processes $X$ and $Y$ evolve according to the system of SDE
\begin{equation}\label{e.cheyette.SDE}
\begin{cases}
\ud X(t)=\big(Y(t)-\kappa(t)\odot X(t)\big)\ud t+\eta(t,X(t))\odot \ud W^\Q(t)\\
\ud Y(t)=\big(\eta^{\odot 2}(t,X(t))-2\kappa(t)\odot Y(t)\big)\ud t
\end{cases}
\end{equation}
subject to the initial conditions $X_{0}=Y_{0}=0$, where we have introduced the quantities $\kappa, \eta\in \R^d$ defined by
\begin{equation}
\kappa_i(t):=-\frac{g_i'(t)}{g_i(t)} \quad \text{and} \quad \eta_i(t,X_i(t)):=h_i(t,X_i(t))g_i(t).
\end{equation}
The following result, whose proof is a direct computation leveraging \eqref{e.p.to.f} and \eqref{e.cheyette.f} shows that modeling the factor processes $X$ and $Y$ is equivalent to modeling zero-coupon bonds. By the \emph{Cheyette model} we will therefore mean the stochastic evolution \eqref{e.cheyette.SDE} of the factor processes $X$ and $Y$.

\begin{proposition}\label{p.zcb.value}
The value of a zero-coupon bond with maturity $T$ at time $t$ in the Cheyette model is given by
\begin{equation}
P(t,T)=\frac{P(0,T)}{P(0,t)}\exp\bigg(-X(t)\cdot G(t,T)-\frac{1}{2}Y(t)\cdot G^{\odot 2}(t,T)\bigg)
\end{equation}
for the deterministic vector-valued function $G(t,T)=(G_i(t,T))_{i\leq d}$ defined by
\begin{equation}\label{e.G.function}
G_i(t,T)=\int_t^T \exp\bigg(-\int_t^s\kappa_i(u)\ud u\bigg)\ud s.
\end{equation}
\end{proposition}
We now discuss how this result can be leveraged to price European and Bermudan Swaptions in the Cheyette model.

\subsection{Pricing European Swaptions in the Cheyette model}

A straightforward no-arbitrage argument reveals that a European Swaption with tenor structure $T_0<T_1<\ldots<T_n$ and fixed rate $K$ is a contingent claim with payoff at the first tenor date $T_0$ given by
\begin{equation}\label{e.european.payoff}
\phi^{\mathrm{EUR}}(X(T_0),Y(T_0)):=\big(1-P(T_0,T_n)-A_{T_0}K\big)_+ \quad \text{where} \quad A_{T_0}:=\sum_{m=1}^n P(T_0,T_m)\Delta T_k
\end{equation}
is the \emph{annuity process} evaluated at the maturity $T_0$ of the Swaption. On the other hand, a direct application of the Feynmann-Kac theorem shows that, in the context of the Cheyette model, the value process 
\begin{equation}
V_t:=V(t,X(t),Y(t))
\end{equation}
of a contingent claim with payoff $V_T:=\phi(X(T),Y(T))$ at its maturity $T>0$ evolves according to the stochastic dynamics
\begin{equation}\label{e.cheyette.BSDE}
\ud V_t = r_t V_t \ud t +\nabla_x V(t,X(t),Y(t))\eta(t,X(t))\odot \ud W^\Q(t)
\end{equation}
subject to the terminal condition $V_T:=\phi(X(T),Y(T))$. Together with Proposition \ref{p.zcb.value} and the observation from \eqref{e.cheyette.f} that
\begin{equation}\label{e.cheyette.r}
r_t=f(0,t)+\sum_{i\leq d}X_i(t),
\end{equation}
this insight will allow us to determine the value of a European Swaption using Neural Networks. More details will be provided in Section \ref{sec:nn_european}.

\subsection{Pricing Bermudan Swaptions in the Cheyette model}

Pricing a Bermudan Swaption requires more care. A no-arbitrage argument analogous to that for a European Swaption shows that the \emph{exercise value}, or the immediate profit, made by exercising the Bermudan Swaption at exercise time $T_m$ is
\begin{equation}\label{e.bermudan.payoff}
\phi^{\mathrm{BER},m}(X(T_m),Y(T_m)):= \big(1-P(T_m,T_n)-A_{T_m}K\big)_+ \quad \text{where} \quad A_{T_m}:=\sum_{\ell=m+1}^n P(T_m,T_\ell)\Delta T_\ell
\end{equation}
is the annuity process evaluated at the exercise date $T_m$. To deduce from this the value $V_{T_m}$ of the Swaption at each exercise date $T_m$, and hence determine its initial value $V_0$, the idea will be to proceed iteratively and backwards in time. More specifically, the strategy will be to compare the exercise value of the Swaption to its \emph{continuation value} at each exercise date $T_m$; the larger of the two being the value of the Swaption at that time. To begin with, notice that the terminal value of the Bermudan Swaption is
\begin{equation}
V_{T_n}:=\phi^{\mathrm{BER},n}(X(T_n),Y(T_n)),
\end{equation}
and observe that the continuation value $C^m=(C_t^m)_{t\in [T_{m-1},T_m)}$ of the Bermudan Swaption between exercise times $T_{m-1}$ and $T_m$ coincides with the value of a contingent claim having payoff $V_{T_m}$ at time $T_m$. It therefore evolves according to the stochastic dynamics
\begin{equation}\label{e.bermudan.continuation}
\ud C_t^m = r_t C_t^m\ud t+\nabla_x C_t^m(t,X(t),Y(t))\eta(t,X(t))\odot \ud W^\Q(t) \quad \text{for } t\in [T_{m-1},T_m)
\end{equation}
subject to the terminal condition $C^m_{T_m}:=V_{T_m}$. With the continuation value $C^m_{T_{m-1}}$ at hand, the value $V_{T_{m-1}}$ of the Bermudan Swaption at exercise time $T_{m-1}$ becomes
\begin{equation}\label{e.bermudan.m.value}
V_{T_{m-1}}:=\max\big(C^m_{T_{m-1}}, \phi^{\mathrm{BER},m-1}\big).
\end{equation}
This well-defined backwards iterative procedure yields the value of the Bermudan Swaption at each tenor date $T_m$. The initial value $V_0$ of the Bermudan Swaption is now the continuation value $C^0_0$ obtained by evolving the SDE \eqref{e.bermudan.continuation} backwards in time from $T_0$ to $0$ subject to the terminal condition $C^0_{T_0}:=V_{T_0}$. This backward iterative procedure will be implemented using a sequence of Neural Networks as detailed in Section \ref{sec:nn_bermudan}.

\section{Neural Networks for Swaption Pricing}\label{sec:NN_for_swaptions}

In this section we describe a Neural Network approach to price European and Bermudan Swaptions in the Cheyette stochastic interest rate model. The strategy will be to define Neural Networks that learn the solution to the SDE \eqref{e.cheyette.BSDE} with an appropriate terminal condition at all points along a partition of the relevant time interval. We also discuss how these Neural Networks can be tensorized using quantum-inspired ideas to yield Tensor Neural Networks.

\subsection{A Neural Network for the European Swaption}
\label{sec:nn_european}

To price a European Swaption with tenor structure $T_0<T_1<\ldots<T_n$ and fixed rate $K\geq 0$, we will use a single Neural Network that learns the solution $V$ to the SDE \eqref{e.cheyette.BSDE} along a partition of the interval $[0,T_0]$ subject to the terminal condition $\phi^{\mathrm{EUR}}$ defined in \eqref{e.european.payoff}. To be more specific, we partition the interval $[0,T_n]$ into $N$ sub-intervals of width $\Delta t = T_n/N$ with endpoints at $0=:t_0<t_1<\ldots<t_N:=T_n$, and we strive to learn the value $V_{t_k}$ of the solution to \eqref{e.cheyette.BSDE} at each of the partition points $t_k$ in the interval $[0,T_0]$. For simplicity, we will assume that $N$ is chosen in such a way that $T_0=t_{k_0}$ for some $0\leq k_0\leq N$; otherwise, in everything that follows, $T_0$ would have to be replaced by its nearest neighbor $t_{k_0}$ in the partition of $[0,T_n]$. Writing $M$ for the batch size, we proceed in $3$ steps.

\subsubsection{The forward simulation}
\label{sec:forward_simulation}
To begin with, we discretize the system of SDE \eqref{e.cheyette.SDE} describing the evolution of the factor processes $X$ and $Y$ using an Euler scheme. This yields $M$ simulated paths $(X^j)_{j\leq M}$ and $(Y^j)_{j\leq M}$ of each factor process. The $k^{\mathrm{th}}$ coordinate in the vectors $X^j=(X^{j,k})_{k\leq N}$ and $Y^j=(Y^{j,k})_{k\leq N}$ is found from the system of discrete differences
\begin{equation}\label{e.cheyette.SDE.discrete}
\begin{cases}
X^{j,k+1} &:= X^{j,k} + \big(Y^{j,k}-\kappa(t_k)\odot X^{j,k}\big)\Delta t +\eta(t_k,X^{j,k})\odot \Delta W^{j,k}\\
Y^{j,k+1} &:= Y^{j,k} +\big(\eta^{\odot 2}(t_k,X^{j,k})-2\kappa(t_k)\odot Y^{j,k}\big)\Delta t
\end{cases}
\end{equation}
subject to the initial condition $X^{j,0}=Y^{j,0}=0$, and it corresponds to a sample of the random variable $X_{t_k}$ or $Y_{t_k}$. Here $\Delta W^{j,k}:= \sqrt{\Delta t} Z^{j,k}$ for a family of independent and identically distributed standard Gaussian random variables $(Z^{j,k})$. Notice that the vectors $Y^j$ are all identical since $Y$ is deterministic; nonetheless, we artificially take $M$ of them as this will lighten the notation required to describe the Neural Network.

\subsubsection{The European terminal condition}

The forward simulation yields $NM$ vectors of triples $(X^{j,k},Y^{j,k},t_k)$ which may be combined with Proposition \ref{p.zcb.value} to estimate the zero-coupon bond prices $P(T_0,T_m)$ required to determine the terminal condition $\phi^{\mathrm{EUR}}$ defined in \eqref{e.european.payoff}. The $m^{\mathrm{th}}$ coordinate in the vector $P^j=(P^{j,m})_{1\leq m\leq n}$ estimates the zero-coupon bond price $P(T_0,T_m)$, and is defined by 
\begin{equation}
P^{j,m}:=\frac{P(0,T_m)}{P(0,T_0)}\exp\bigg(-X^{j,k_0}\cdot G(T_0,T_m)-\frac{1}{2}Y^{j,k_0}\cdot G^{\odot 2}(T_0,T_m)\bigg),
\end{equation}
where the deterministic function $G$ is given by \eqref{e.G.function} and $k_0$ is the unique index with $T_0=t_{k_0}$. From these zero-coupon bond estimates, it is possible to approximate the $M$ terminal conditions
\begin{equation}
\phi^{\mathrm{EUR},j}:=\bigg(1-P^{j,n}-\sum_{m=1}^n KP^{j,m}\Delta T_m\bigg)_+
\end{equation}
required to price the European Swaption.

\subsubsection{The European Neural Network}
\label{sec:european_NN}
The $NM$ vectors of triples $(X^{j,k},Y^{j,k},t_k)$ and the $M$ terminal conditions $(\phi^{\mathrm{EUR},j})$ now become the inputs of the Neural Network tasked with learning the solution to the SDE \eqref{e.cheyette.BSDE} subject to the terminal condition $\phi^{\mathrm{EUR}}$ at the points $t_k$ in the interval $[0,T_0]$. The output of this Neural Network are $M$ paths
\begin{equation}
\widehat{V}^j(\theta)=\big(\widehat{V}^{j,k}(\theta)\big)_{k\leq k_0}
\end{equation}
with $\widehat{V}^{j,k}(\theta)$ being an estimator for $V_{t_k}$. This output vector can be used to estimate the gradient $\widehat{\nabla}_{X} V^{j,k}$ by means of automatic differentiation, and this estimated gradient can be leveraged to obtain another approximation of $V_{t_k}$ through an Euler discretization of the backward SDE \eqref{e.cheyette.BSDE},
\begin{equation}
\widetilde{V}^{j,k+1}(\theta):=\widehat{V}^{j,k}(\theta) +r^{j,k}\widehat{V}^{j,k}(\theta)\Delta t +\widehat{\nabla}_XV^{j,k}\eta(t_k,X^{j,k})\odot \Delta W^{j,k},
\end{equation}
where
\begin{equation}\label{e.approximate.r}
r^{j,k}:=f(0,t_k)+\sum_{i\leq d}X^{j,k}
\end{equation}
is an approximation of the short rate \eqref{e.cheyette.r} at $t_k$. The Neural Network is now trained by minimizing the square difference between the estimators $\widehat{V}^j$ and $\widetilde{V}^j$ while matching the terminal conditions $\phi^{\mathrm{EUR},j}$ across batches. In other words, its associated loss function is defined by
\begin{equation}
\mathcal{L}(\theta):=\sum_{j=1}^M \sum_{k=1}^N \big(\widehat{V}^{j,k}(\theta)-\widetilde{V}^{j,k}(\theta)\big)^2+\sum_{j=1}^M \big(\widehat{V}^{j,k_0}(\theta)-\phi^{\mathrm{EUR},j}\big)^2
\end{equation}
Notice that we could have included a term comparing $\widetilde{V}_N^{j,k_0}(\theta)$ and $\phi^{\mathrm{EUR},j}$, but this is not necessary as the triangle inequality already takes care of it.
\begin{figure}[h]
    \centering
    \includegraphics[scale = 0.55]{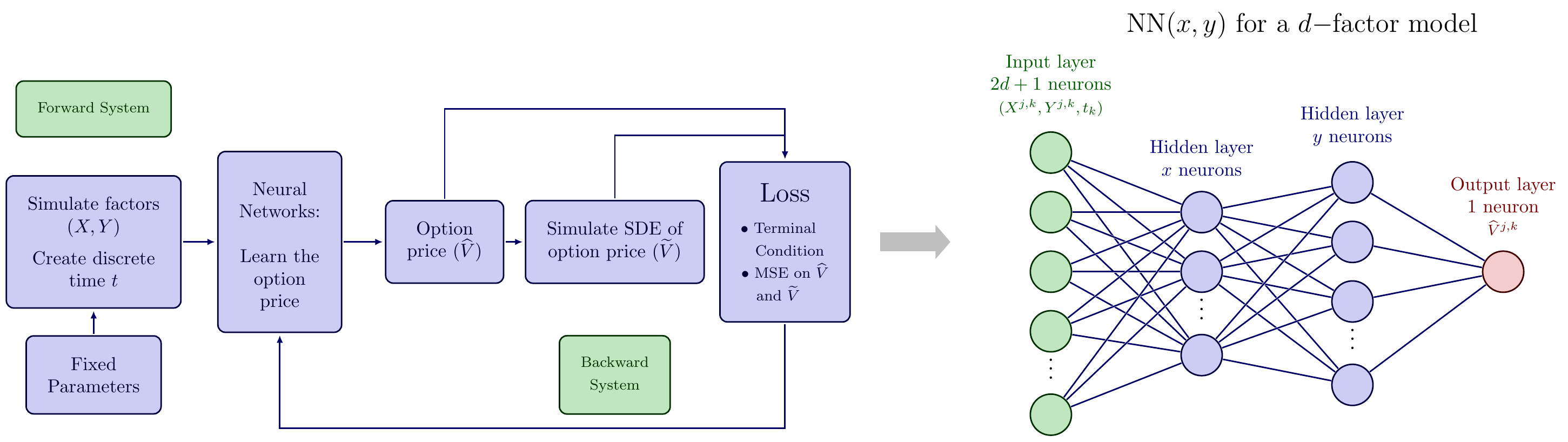}
    \caption{Learning pipeline for the European Swaption Neural Network.}
    \label{Eur}
\end{figure}

\subsection{A Sequence of Neural Networks for the Bermudian Swaption}
\label{sec:nn_bermudan}

To price a Bermudan Swaption with tenor structure $T_0<T_1<\ldots<T_n$ and fixed rate $K\geq 0$, we will use a backward sequence of $n+1$ Neural Networks. The $m^{\mathrm{th}}$ of these Neural Networks will learn the solution $C^m$ to the SDE \eqref{e.bermudan.continuation} along a partition of the interval $[T_{m-1},T_m]$ subject to a terminal condition depending on the payoff $\phi^{\mathrm{BER},m}$ defined in \eqref{e.bermudan.payoff} and the ouput of the previous Neural Network. Here and henceforth we adopt the convention that $T_{-1}:=0$. Just like we did for European Swaptions, we partition the interval $[0,T_n]$ into $N$ sub-intervals of width $\Delta t = T_n/N$ with endpoints at $0=:t_0<t_1<\ldots<t_N:=T_n$, and we assume that $N$ is chosen in such a way that for every $0\leq m\leq n$, we have $T_m=t_{k_m}$ for some $0\leq k_m\leq N$; otherwise, in everything that follows, $T_m$ would have to be replaced by its nearest neighbour $t_{k_m}$ in the partition of $[0,T_n]$. Moving forward we adopt the convention that $t_{k_{-1}}:=0$. Writing $M$ for the batch size we once again proceed in $3$ steps.

\subsubsection{The forward simulation}

The forward simulation is identical to that described in Section \ref{sec:forward_simulation} for European Swaptions, and again yields $NM$ vectors of triples $(X^{j,k},Y^{j,k},t_k)$.

\subsubsection{The Bermudan payoffs}

To price a Bermudan Swaption, we need to estimate the $n+1$ payoffs $\phi^{\mathrm{BER},m}$ defined in \eqref{e.bermudan.payoff}. To do this we require approximations $P^{j,m,\ell}$ to the zero-coupon bond prices $P(T_m,T_\ell)$ for $0\leq m< n$ and $m<\ell\leq n$. These may be obtained from Proposition \ref{p.zcb.value} by setting
\begin{equation}
P^{j,m,\ell}:=\frac{P(0,T_\ell)}{P(0,T_m)}\exp\bigg(-X^{j,k_m}\cdot G(T_m,T_\ell)-\frac{1}{2}Y^{j,k_m}\cdot G^{\odot 2}(T_m,T_\ell)\bigg),
\end{equation}
where the deterministic function $G$ is given by \eqref{e.G.function} and $k_m$ is the unique index with $T_m=t_{k_m}$. From these zero-coupon bond estimates, for each of the $M$ batches, it is possible to approximate the $m^{\mathrm{th}}$ payoff
\begin{equation}
\phi^{\mathrm{BER},j,m}:= \bigg(1-P^{j,m,n}-\sum_{\ell=m+1}^n KP^{j,m,\ell}\Delta T_\ell\bigg)_+ 
\end{equation}
required to price the Bermudan Swaption.

\subsubsection{The sequence of Bermudan Neural Networks}

The $NM$ vectors of triples $(X^{j,k},Y^{j,k},t_k)$ with $k_{n-1}\leq k\leq k_n$ and the $M$ payoffs $(\phi^{\mathrm{BER},j,n})$ now become the inputs of the $n^{\mathrm{th}}$ Neural Network tasked with learning the solution to the SDE \eqref{e.bermudan.continuation} subject to the terminal condition $\phi^{\mathrm{BER},n}$ at the points $t_k$ in the interval $[T_{n-1},T_n]$. The output of this Neural Network are $M$ continuation values
\begin{equation}
\widehat{C}^{j,n}(\theta_n)=\big(\widehat{C}^{j,n,k}(\theta_n)\big)_{k_{n-1}\leq k\leq k_n}
\end{equation}
with $\smash{\widehat{C}^{j,n,k}(\theta_n)}$ being an estimator for $\smash{C^n_{t_k}}$. Before discussing how the $m^{\mathrm{th}}$ Neural Network learns, let us describe the inputs and outputs of the $m^{\mathrm{th}}$ Neural Network in this stacked sequence of Neural Networks for $0\leq m< n$. The $NM$ vectors of triples $(X^{j,k}, Y^{j,k},t_k)$ with $k_{m-1}\leq k\leq k_m$ and the $M$ terminal conditions
\begin{equation}
V^{j,m} := \max\big(\phi^{\mathrm{BER},j,m}, \widehat{C}^{j,m+1,k_{m}}(\theta_{m+1})\big)
\end{equation}
are the inputs of the $m^{\mathrm{th}}$ Neural Network tasked with learning the solution to the SDE \eqref{e.bermudan.continuation} subject to the terminal condition $V_{T_m}$ defined in \eqref{e.bermudan.m.value} at the points $t_k$ in the interval $[T_{m-1},T_m]$. The output of the $m^{\mathrm{th}}$ Neural Network are the $M$ continuation values
\begin{equation}
\widehat{C}^{j,m}(\theta_m)=\big(\widehat{C}^{j,m,k}(\theta_m)\big)_{k_{m-1}\leq k\leq k_m}
\end{equation}
with $\smash{\widehat{C}^{j,m,k}(\theta_m)}$ being an estimator for $\smash{C^m_{t_k}}$. The value of the Bermudan swaption is then obtained from the continuation values $V_0^j:=\widehat{C}^{j,0,0}(\theta_0)$.

Having described the stacked Neural Network structure, let us delve into the learning procedure for the $m^{\mathrm{th}}$ Neural Network. This will be identical for $0\leq m\leq n$ and will very closely resemble the procedure described for the European Neural Network in Section \ref{sec:european_NN}. The output $\smash{\widehat{C}^{j,m,k}(\theta_m)}$ of the $\smash{m^{\mathrm{th}}}$ Neural Network can be used to estimate the gradient $\smash{\widehat{\nabla}_X C^{j,m,k}}$ by means of automatic differentiation, and this estimated gradient can be leveraged to obtain another approximation of the continuation value $C^m_{t_k}$ through the Euler discretization of the backward SDE \eqref{e.bermudan.continuation},
\begin{equation}
\widetilde{C}^{j,m,k+1}(\theta_m) = \widehat{C}^{j,m,k}(\theta_m)+r^{j,k} \widehat{C}^{j,m,k}(\theta_m)\Delta t+\widehat{\nabla}_X C^{j,m,k}\eta(t_k,X^{j,k})\odot \Delta W^{j,k},
\end{equation}
where $r^{j,k}$ is an approximation of the short rate \eqref{e.cheyette.r} at $t_k$. The $m^{\mathrm{th}}$ Neural Network is now trained by minimizing the loss function 
\begin{equation}
\mathcal{L}_m(\theta_m):=\sum_{j=1}^M \sum_{k=k_{m-1}}^{k_m} \big(\widehat{C}^{j,m,k}(\theta_m)-\widetilde{C}^{j,m,k}(\theta_m)\big)^2+\sum_{j=1}^M \big(\widehat{C}^{j,m,k_m}(\theta_m)-V^{j,m}\big)^2
\end{equation}
which strives to match the estimators $\widehat{C}^{j,m}$ and $\widetilde{C}^{j,m}$ as well as the terminal conditions $V^{j,m}$ across batches.

\begin{figure}[!htp]
    \centering
    \includegraphics[scale = 0.47]{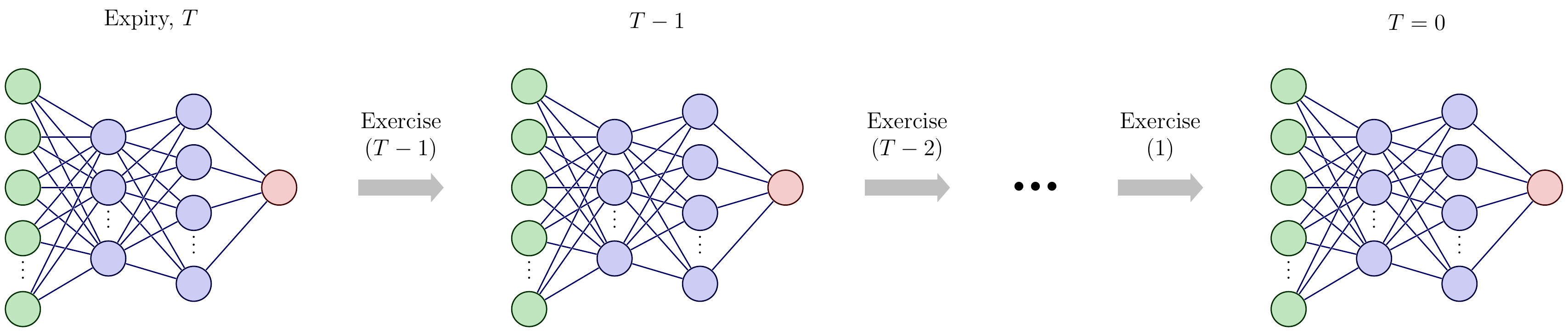}
    \caption{Learning pipeline for the Bermudan Swaption stacked Neural Network.}
    \label{Ber}
\end{figure}

\section{Tensorizing the Neural Networks}\label{sec:TNN}

\begin{figure}[h]
\centering
\includegraphics[scale=0.9]{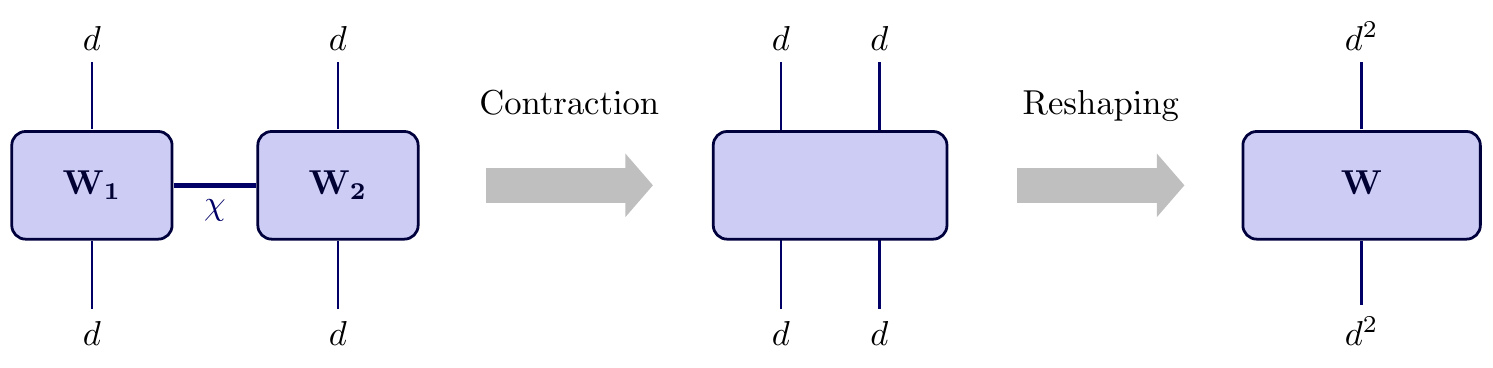}
\caption{The process of contracting a 2-node MPO and reshaping it into the weight matrix $\mathbf{W}$ in each forward pass.}
\label{Fig:Fig3}
\end{figure}

To address the shortcomings of dense architectures, we transform classical fully-connected Dense Neural Networks into what we call Tensor Neural Networks. This has the purpose of enhancing training performance and reducing memory consumption \cite{CACIB_1, patel2022quantuminspired}. The way we tensorize the Neural Networks is to replace the weight matrix of a dense layer by a Tensor Network \cite{RomanTN}. In particular, we choose a Matrix Product Operator (MPO) representation \cite{RomanTN} of the weight matrix that is analogous to the Tensor-Train format \cite{TNN_NIPS}, and we call this layer a \emph{TN layer}. This representation, however, is not unique, and is determined by two additional parameters: the MPO bond dimension, and the number of tensors in the MPO. In the simplest case, the MPO may consist of only two tensors, $\mathbf{W_1}$ and $\mathbf{W_2}$, as shown in Figure \ref{Fig:Fig3}. The MPO in the figure has bond dimension $\chi$ and physical dimension $d$ as the input and output dimension. The TN layer with such an MPO can be initialized in the same manner as a weight matrix of a dense layer.\newline

In the forward pass of the TN layer, we first contract the MPO along the bond index and then reshape the resulting rank-4 tensor into a matrix as shown in Figure \ref{Fig:Fig3}. This matrix is the corresponding weight matrix of a TN layer. The weight matrix can then be multiplied with the input vector. We apply an activation function to the resulting output vector, thereby finishing the forward pass. The weight matrix takes the form 
\begin{equation}
    \mathbf{W} = \sum_{\alpha = 1}^{\chi} (\mathbf{A}_\alpha \otimes \mathbf{B}_\alpha), ~~~\mathbf{W} \in \mathbb{R}^{d^2 \times d^2},
\label{sum_of_tensor_product}
\end{equation}
where $\mathbf{W}_1 = [\mathbf{A}_1, \mathbf{A}_2, \cdots, \mathbf{A}_{\chi}], \mathbf{A}_\alpha \in \mathbb{R}^{d \times d}$ and $\mathbf{W}_2 = [\mathbf{B}_1, \mathbf{B}_2, \cdots, \mathbf{B}_{\chi}], \mathbf{B}_\alpha \in \mathbb{R}^{d \times d}$ are the two rank-3 weight tensors connected by a virtual bond $\alpha$ of dimension $\chi$. The resulting weight matrix $\mathbf{W}$ is of dimension $d^2 \times d^2$, so it contains $d^4$ elements. Notice that these elements \emph{are not independent} due to the Kronecker product in \eqref{sum_of_tensor_product}. Indeed, the weights come from the TN structure with $2 \chi d^2$ trainable parameters, and they are a sum of products of elements of the tensors $\mathbf{A}$ and $\mathbf{B}$, thus leading to a correlation structure. It is worth noting that the parameters to be trained are not the matrix elements of the weight matrix, but the elements of the individual tensors of the MPO. This can lead to interesting training behaviour and can result in faster convergence of the loss function. Moreover, any choice of $\chi < d^2 / 2$ will result in $d^4 - 2 \chi d^2$ fewer trainable parameters than an equivalent dense layer, thus allowing for potential parameter savings. In principle, when $\chi = d^2$, we have sufficient degrees of freedom to be able to construct an arbitrary $d^2 \times d^2$ matrix. By increasing the bond dimension, we therefore expect the TN layer to behave increasingly like a dense layer \cite{CACIB_1}.

\section{Swaption Pricing Results}\label{sec:results}

We now apply the methodology described in Sections \ref{sec:nn_european}, \ref{sec:nn_bermudan} and \ref{sec:TNN} to benchmark the performance of Dense Neural Networks and Tensor Neural Networks in pricing European and Bermudan Swaptions. For benchmarking, we consider the Cheyette model with constant parameters and identity correlation matrix. Under this parametrization, the Cheyette model is equivalent to a one-factor Hull and White model. It is worth noting, however, that our approach and code can be extended to the generalized version of the Cheyette model.

\subsection{European Swaptions}

To price European Swaptions we place ourselves in the context of the 3-factor Cheyette model with $X_i(0) = Y_i(0)=0$ for $1\leq i \leq 3$, we use the base parameters $\kappa = -0.02$ and $\eta = 0.0065$, the tenor structure $(T_0,T_1,T_2,T_3,T_4) = (1,2,3,4,5)$, the fixed rates $K=0.00$ or $K=0.01$, and the initial forward curve $T\mapsto f(0,T)$ implied from \eqref{e.f.from.P} and the zero-coupon bond prices $P(0,T)$ given in Table \ref{table:discount curve}. Moreover, we partition the time interval $[0,T_4]$ into $N = 500$ sub-intervals of width $\Delta t = 0.01$. 
\begin{table}[h]
\centering
\begin{tabular}{|c |c || c|c|| c|c|}
\hline
Maturity  & Bond price  & Maturity & Bond price  & Maturity & Bond price \\
\hline
0  & 1.00000 & 5  & 0.88232 & 18 & 0.60911 \\
1  & 0.99005 & 6  & 0.83500 & 23 & 0.53693 \\
2  & 0.97528 & 7  & 0.78240 & 28 & 0.49611 \\
3  & 0.95596 & 8  & 0.77064 & 33 & 0.47940 \\
4  & 0.91376 & 13 & 0.67661 & 38 & 0.46721 \\
\hline
\end{tabular}
\caption{Zero-coupon bond prices $T\mapsto P(0,T)$ used to determine the initial forward curve $T\mapsto f(0,T)$.}
\label{table:discount curve}
\end{table}
For our use-case we only use Neural Networks with a 2-hidden layer architecture. Furthermore, for Tensor Neural Networks, we only construct TN layers that are symmetric in each input and each output dimension. As a result, we choose the first layer in our Tensor Neural Network to be a dense layer with neurons that match the input shape of the second TN layer. For convenience of notation, we will write DNN$(x,y)$ to mean a two-layer Dense Neural Network with $x$ neurons in the first layer and $y$ neurons in the second layer. Similarly, we will write TNN$(x, y)$ to mean a two-layer Tensor Neural Network with a dense first layer having $x$ neurons and a TN second layer having $y$ neurons, where we must have $x=y$. To optimize the Neural Network weights, we use the Adam Optimizer with batch of size 100 and a piece-wise learning rate of $[10^{-2}, 10^{-3}, 10^{-4}, 10^{-5}]$. This means that for the $i$'th quarter of epochs, we use a learning rate of $10^{-(i+1)}$ in the Adam Optimizer.

Having described the model and Neural Network specifications for our numerical experiments, we now turn our attention to the numerical results for the initial price and loss evolution of European Swaptions with fixed rate $K=0.00$ and $K=0.01$. We focus on three architectures, TNN(64, 64), DNN(64, 64) and DNN(24,27), and we compare and contrast them to the Monte-Carlo (MC) price obtained by approximating the discounted payoff
\begin{equation}
\E\bigg[\exp\bigg(-\int_0^{T_0} r_s\ud s\bigg) \phi^{\mathrm{EUR}}(X(T_0),Y(T_0))\bigg]
\end{equation}
by its sample average. In the context of European Swaptions, we will assume this MC price to be the ``true'' option price. We start with DNN(64, 64) as it is the smallest Dense Neural Network that agrees with the MC price for one of the two fixed rates $K$ that we consider. To compare a Tensor Neural Network with this DNN(64, 64), we select the architecture TNN(64, 64) as it has a comparable number of neurons. However, it is worth noting that TNN(64, 64) does not have the same number of parameters as DNN(64, 64). In our experiments, we use a bond-dimension $\chi$= 2, so, as discussed in Section \ref{sec:TNN}, the architecture TNN(64,64) would have $d^4 - 2\chi d^2 = (8)^4 - 2*2*(8^2) = 3840$ fewer parameters than DNN(64,64). Notice that $d=8$ as $64=8^2$ (see Figure \ref{Fig:Fig3}). In this particular case, DNN(64, 64) has $4737$ parameters whereas TNN(64, 64) has $897$ parameters. Moreover, to have a fair comparison, we also contrast TNN(64, 64) to DNN(24, 27) as it is the best performing Dense Neural Network architecture with a comparable number of parameters as TNN(64,64).
\newpage
\begin{figure}[!htp]
    \centering
    \includegraphics[scale = 0.77]{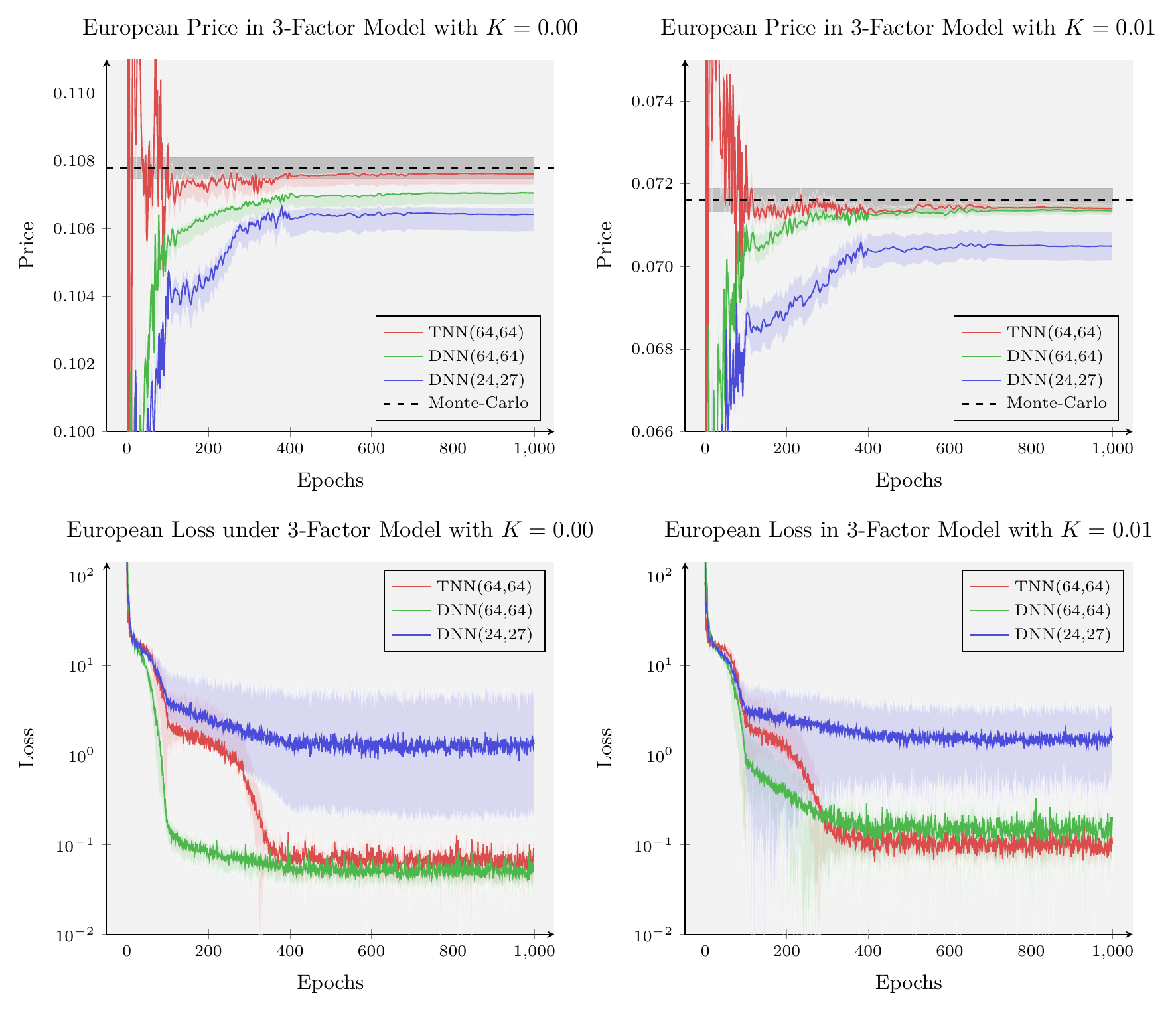}
    \caption{(\textbf{top panel}) Initial option price evolution for fixed rate $K = 0.00$ (left) and fixed rate $K=0.01$ (right) for TNN(64, 64) with a bond dimension 2 (red), the corresponding DNN(64,64) with similar neuron count (green) and the best DNN(24,27) with equivalent parameter count (blue). The plots display the mean and $95\%$ confidence interval for the runs. To benchmark the results, the dotted line (black) indicates the MC price from $10^5$ runs while the grey shaded region indicates its associated $95\%$ confidence interval. (\textbf{bottom panel}) Training loss evolution for fixed rate $K=0.00$ (left) and fixed rate $K=0.01$ (right) for TNN(64, 64) with a bond dimension 2 (red), the corresponding DNN(64,64) with similar neuron count (green) and the best DNN(24,27) with equivalent parameter count (blue). The plots display the mean and $95\%$ confidence interval for the runs.}
    \label{Fig: Eur_2_cols}
\end{figure}

The results in Figure \ref{Fig: Eur_2_cols} show that the Tensor Neural Network TNN(64, 64) outperforms the best performing Dense Neural Network with similar parameter count for both fixed rates $K$ considered. This is evident from the option price evolution, with the Tensor Neural Network approaching the MC price faster and to a much greater extent while exhibiting a significantly smaller confidence interval than the DNN(24, 27) architecture which doesn't even converge to the MC price. This is supported by the loss function evolution, with the Tensor Neural Network loss function becoming considerably smaller than the DNN(24,27) loss function. Upon comparing the TNN(64, 64) architecture to a dense architecture with similar neuron count but significantly higher parameter count, we see that the Tensor Neural Network again outperforms its dense counterpart despite an $81\%$ compression in parameter count. It is worth pointing out that while the DNN(64,64) architecture does result in convergence to the MC price for the fixed rate $K=0.01$, it fails to converge to this benchmark price for the fixed rate $K=0.00$. On the other hand, the Tensor Neural Network readily converges to this ``true'' price for both fixed rates. In summary, the Tensor Neural Network outperforms both the best Dense Neural Network with comparable number of parameters and the Dense Neural Network with similar neuron count but considerably more parameters.

\begin{figure}[!htp]
    \centering
    \includegraphics[scale = 0.77]{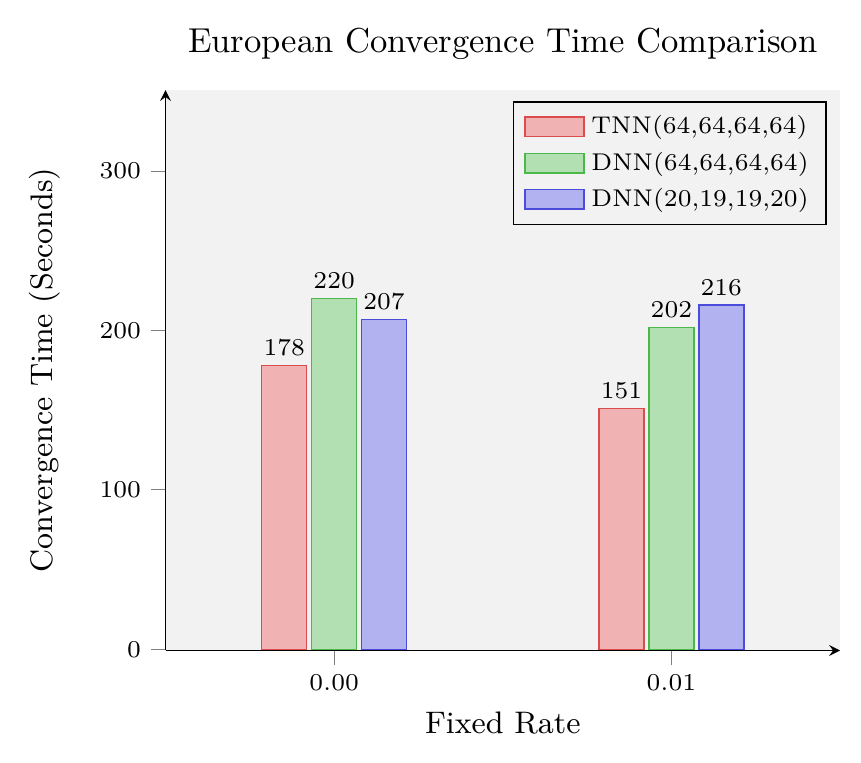}
    \caption{European Swaption Price for different architectures in 3-factor Cheyette Model with fixed rate $K=0.00$.}
    \label{Fig: Eur_convergence_speed_plot}
\end{figure}

In light of the fact that the two layer Dense Neural Networks in Figure \ref{Fig: Eur_2_cols} fail to converge, to compare the convergence time of Tensor Neural Networks and Dense Neural Networks we consider four layer architectures. The three four layer architectures shown in Figure \ref{Fig: Eur_convergence_speed_plot} all converge to the true price with the Tensor Neural Network doing so considerably faster than its dense counterparts with comparable number of neurons or parameters.
\newpage
\vspace*{-1.2cm}
\vspace*{-0.5cm}
\subsection{Bermudan Swaptions}

To price Bermudan Swaptions we consider the 3-factor Cheyette model with $X_i(0) = Y_i(0)=0$ for $1\leq i \leq 3$, we use the base parameters $\kappa = -0.02$ and $\eta = 0.0065$, the tenor structure $(T_0,T_1,T_2,T_3,T_4) = (1,2,3,4,5)$, the fixed rates $K=0.00$ or $K=0.01$, and the initial forward curve $T\mapsto f(0,T)$ implied from \eqref{e.f.from.P} and the zero-coupon bond prices $P(0,T)$ given in Table \ref{table:discount curve}. Moreover, we partition the time interval $[0,T_4]$ into $N = 500$ sub-intervals of width $\Delta t = 0.01$, and we assume that the exercise dates of the Bermudan Swaption coincide with its tenor structure.
\vspace*{-0.2cm}
\begin{figure}[!htp]
    \centering
    \includegraphics[scale = 0.77]{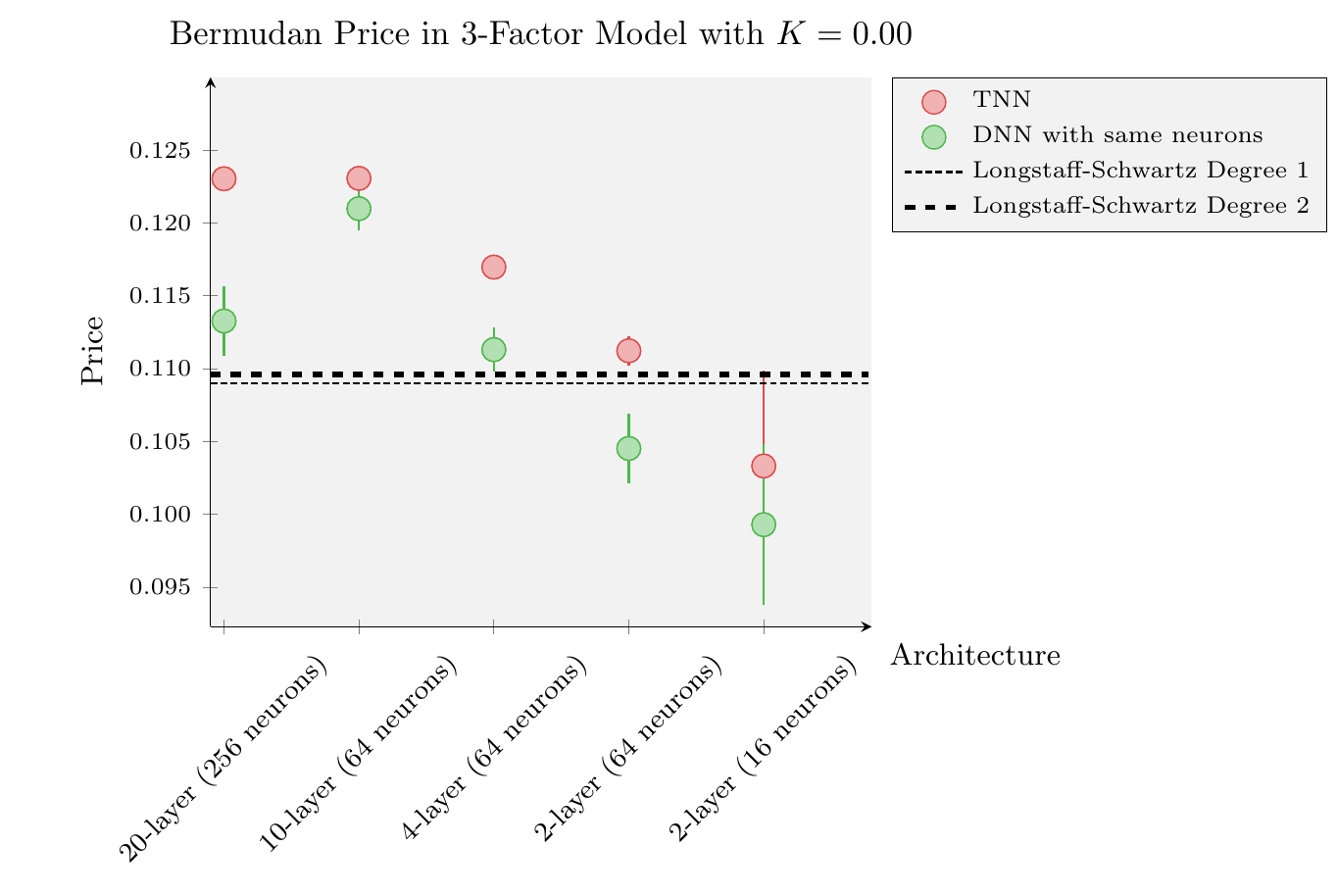}
    \caption{Bermudan Swaption Price for different architectures in 3-factor Cheyette Model with fixed rate $K=0.00$.}
    \label{Fig: Ber_main_result}
\end{figure}
\vspace*{-0.2cm}

The plot in Figure \ref{Fig: Ber_main_result} gives the Bermudan Swaption price for a collection of different Neural Network architectures both for Tensor Neural Networks and Dense Neural Networks. We have chosen to compare a given Tensor Neural Network architecture to the Dense Neural Network architecture with the same number of neurons, and therefore a considerably higher parameter count. The differences in parameter count between comparable Tensor Neural Networks and Dense Neural Networks is summarized in Table \ref{table:parameter count}.

\begin{table}[h]
\centering
\small{
\begin{tabular}{|c|c|c|c|}
\hline
Layers  & Neurons  & Dense Neural Networks  & Tensor Neural Networks \\
\hline
2  & 16 & 417  & 225  \\
2  & 64 & 4,737  & 897  \\
4  & 64 & 13,057  & 1,537  \\
10  & 64 & 38,017  & 3,457  \\
20  & 256 & 1,252,353 & 26,625 \\
\hline
\end{tabular}}
\caption{Parameter Count for Dense Neural Networks and Tense Neural Networks compared in Figure \ref{Fig: Ber_main_result}.}
\label{table:parameter count}
\end{table}
\newpage
To benchmark the Bermudan Swaption prices given by Tensor Neural Networks and Dense Neural Networks, Figure \ref{Fig: Ber_main_result} also displays the Bermudan Swaption price provided by the classical Longstaff-Schwartz (LS) approach. We have included the LS price when the regressors are of degree 1, which turned out to be \$0.1090, and also the price when the regressors are of degree 2, which turned out to be \$0.1096. It is well-known that the LS approach leads to an under-estimate of the option price \cite{Glas04} and that the LS price approaches the ``true'' option price when the degree of the regressors tends to infinity \cite{l-s-algo}. The first of these heuristics is supported by our numerical results, and the second suggests that the ``true'' option price should be higher than the \$0.1096 given by the LS with regressors of degree 2. It is worth noting that such a price will also be higher than the ``true'' price of \$0.1079 for the European Swaption with the same parameters displayed in Figure \ref{Fig: Eur_2_cols}. This is in accordance with the fact that the price of a Bermudan Swaption should always be lower bounded by the price of its European counterpart.

\begin{figure}[!htp]
    \centering
    \includegraphics[scale = 0.77]{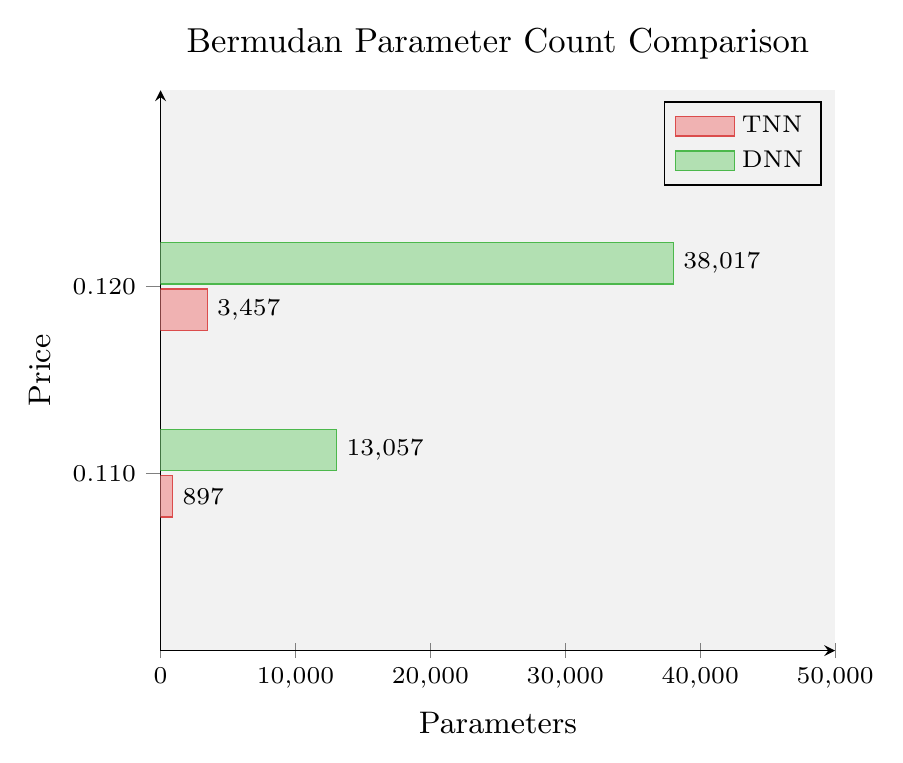}
    \caption{
    DNN and TNN parameter comparison for Bermudan Swaption price in $3$-factor Cheyette model with $K=0.00$ to cross \$0.110 and \$0.120 price threshold .}
    \label{Fig: Ber_parameter_comparison}
\end{figure}

Having understood how we will benchmark the Tensor Neural Network and Dense Neural Network prices, and emphasising that this benchmark is not as reliable as it was in the setting of European Swaptions, let us turn our attention to the Neural Network prices for various architectures. To begin with we used a 2-layer architecture with 16 neurons in each layer for each of the 5 Neural Networks (one for each exercise date), and trained each Neural Network for 500 epochs. The prices for such a Tensor Neural Network and Dense Neural Network are displayed on the far right of Figure \ref{Fig: Ber_main_result}. It is clear that these prices are underestimates resulting from the Neural Networks under-fitting the data. This led us to consider the same configuration of Neural Networks with 64 neurons instead of just 16 in each of the 2 layers. We observe a marked improvement in both the price and the stability (the bars around the estimate indicate a 95\% confidence interval for 10 runs under the same configuration) of the resulting Bermudan Swaption prices, particularly in the Tensor Neural Network setting where the price goes above the LS price as expected of the ``true'' option price. Upon increasing the complexity of the architecture to a 4-layer Neural Network with 64 neurons in each layer, the improvements are evident with both the Tensor Neural Network and the Dense Neural Network above the LS price, thereby making the Neural Network advantage evident. This phenomenon of Neural Networks outperforming classical regression should come as no surprise. Indeed, the finite number of regressors chosen in the classical LS approach cannot fully represent the conditional payoff they strive to approximate. This limitation could be mitigated by increasing the degree of the regressors but this quickly becomes infeasible from an accuracy and time perspective, leading both to over-fitting and a lack of convergence. The advantage of Neural Networks is that they allow us to learn the relevant regressors instead of selecting them a priori, thus being more amenable to high-dimensional settings. Once the Tensor Neural Networks and Dense Neural Networks both start to outperform the LS approach, in the sense that they yield higher prices, we increase the number of layers to 10 and see a further increment in price to \$0.1230. Upon increasing the number of layers to 20, we observe that the Tensor Neural Network price is almost identical to that given in the 10 layer setting. On the other hand, the Dense Neural Network price for 20 layers falls subject to over-fitting and is considerably worse than the price observed in the 10 layer setting both in terms of accuracy and stability. Besides exhibiting overfitting, as shown in Figure \ref{Fig: Ber_parameter_comparison}, Dense Neural Networks also require a larger number of parameters than Tensor Neural Networks to converge above the threshold prices of \$0.110 and \$0.120 which appear to be lower bounds for the true price. This suggests that Tensor Neural Networks provide better and more scalable prices than both LS and Dense Neural Networks which are more prone to overfitting if not properly fine-tuned.

\newpage
Before closing this section we would like to delve deeper into the 150bps difference between the degree 2 LS price and the 20 layer Tensor Neural Network price shown in Figure \ref{Fig: Ber_main_result} as this might seem overly large. We hypothesise that this difference is a consequence of using low degree regressors in the LS approach, and that the LS price would approach the Tensor Neural Network price if we increased the degree of the regressors. Due to computational limitations, instead of increasing the degree of the regressors, we decrease the complexity of the model by considering the 1-factor Cheyette model as opposed to the 3-factor model.

\begin{figure}[!htp]
    \centering
    \includegraphics[scale = 0.77]{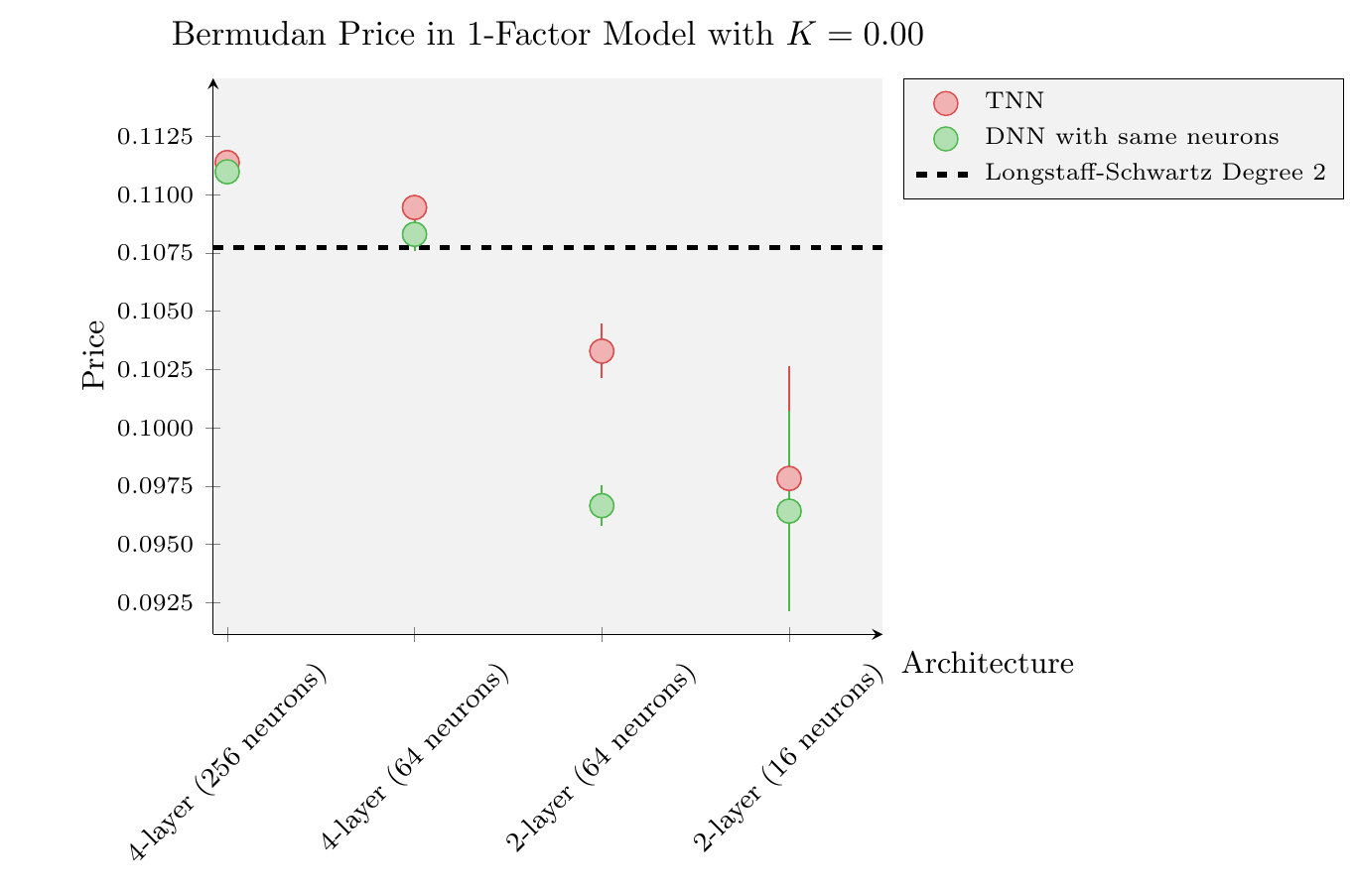}
    \caption{Bermudan Swaption Price for different architectures in 1-factor Cheyette Model with fixed rate $K=0.00$.}
    \label{Fig: Ber_main_result_one_factor}
\end{figure}

The hypothesis that the 150bps difference is being driven by an under-fitting of the regressors is supported by Figure \ref{Fig: Ber_main_result_one_factor} which shows a smaller 30bps difference between the degree 2 LS price and the 4 layer Tensor Neural Network price. Furthermore, in this simpler setting the Tensor Neural Network and the Dense Neural Network price both tend to the same value with almost no standard error. To further display the advantage provided by Tensor Neural Networks, let us compare the loss plots, parameter count and time to convergence for Tensor Neural Networks and Dense Neural Networks. 

\begin{figure}[!htp]
    \centering
    \includegraphics[scale = 0.77]{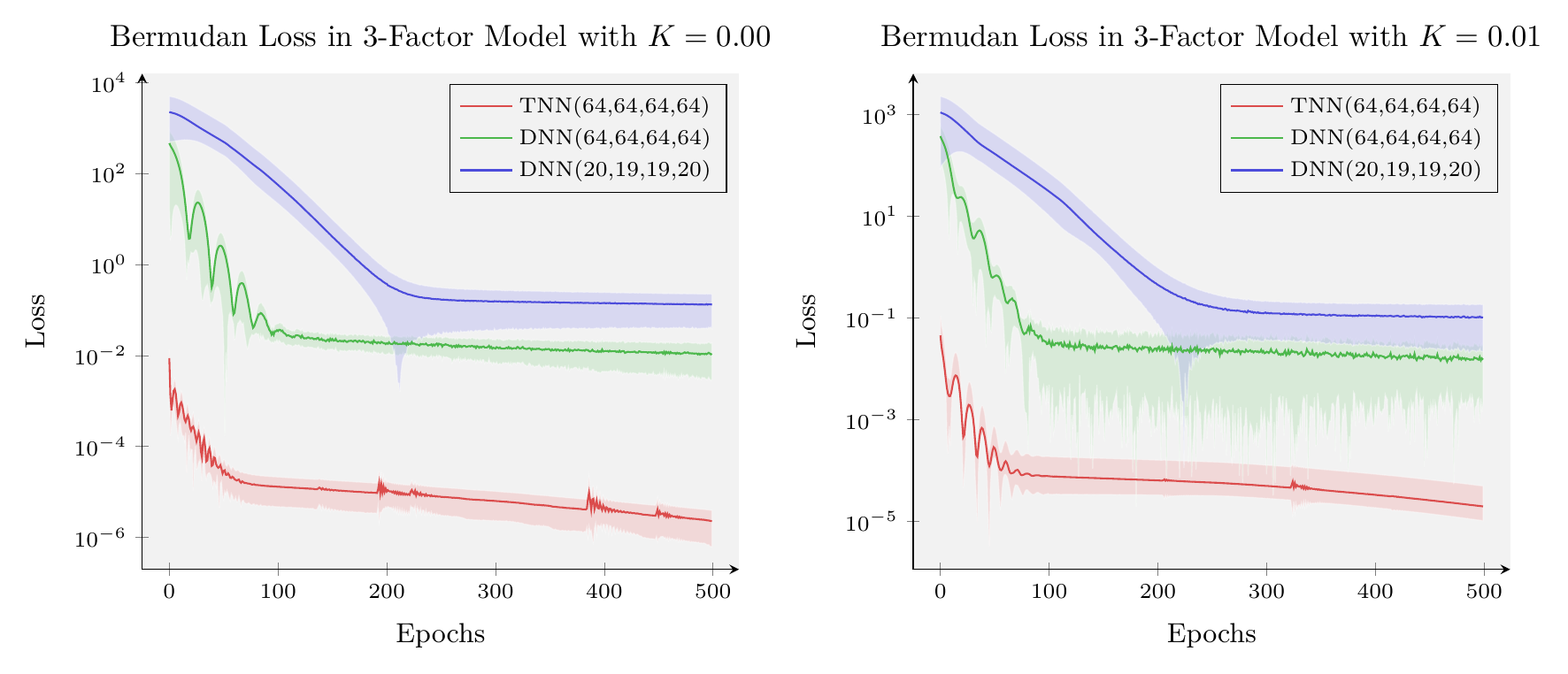}
    \caption{Bermudan Swaption loss evolution in $3$-factor Cheyette model with fixed rates $K=0.00$ and $K=0.01$.}
    \label{Fig: Ber_2_cols}
\end{figure}

From the premise that smaller losses imply better results, Figure \ref{Fig: Ber_2_cols} shows that the Tensor Neural Network outperforms the Dense Neural Network with similar neuron count and also the best Dense Neural Network with comparable parameter count. This corroborates the observation from Figures \ref{Fig: Ber_main_result} and \ref{Fig: Ber_main_result_one_factor} that Tensor Neural Networks provide more reliable prices than Dense Neural Networks. 

\section{Conclusions and Outlook}\label{sec:conclusion}

In this paper we have shown how Tensor Neural Networks can be leveraged to price European Swaptions in the multi-factor Cheyette model. We have also extended this approach to price early-exercise Swaptions under this stochastic interest rate model by stacking $n$ feed-forward Neural Networks to price a Bermudan Swaption with $n$ exercise dates. To quantify the considerable advantages of Tensor Neural Networks we have performed rigorous empirical benchmarking. In doing so, we have demonstrated that Tensor Neural Networks provide significant parameter savings relative to Dense Neural Networks with comparable number of neurons or parameter count while attaining more reliable prices having smaller variance. We have further shown that Tensor Neural Networks achieve a computational speed-up in training when contrasted with Dense Neural Networks having comparable number of neurons or parameter count. Although we have performed numerical experiments in the 3-factor setting, unlike the traditional Monte-Carlo based methods which succumb to the Curse of Dimensionality, our approach can be scaled to a significantly higher number of factors. Despite an absence of theoretical bounds, which can be an area of further investigation, the Tensor Neural Network approach described in this paper can be used to improve training from a memory and speed perspective for a wide variety of problems in Machine Learning. Extending this approach to yield a parametric Tensor Neural Network, able to instantaneously derive prices for a wide variety of different model parameters without having to retrain the Neural Network, would be an interesting and fruitful avenue of further investigation.

\bigskip 
\noindent
{\bf Acknowledgements -} This work was supported by the European Union’s Horizon Europe research and innovation programme under Grant Agreement No 190145380. We would also like to acknowledge the regular fruitful discussions with the technical teams both at BBVA and Multiverse Computing. 

\bigskip 
\noindent
{\bf Disclaimer -} This paper is purely scientific and informative in nature and is not a product of BBVA SA or any of its subsidiaries. Neither BBVA nor such subsidiaries are aware of or
necessarily share the premises, conclusions or contents in general of this document.
Consequently, the responsibility for its originality, accuracy, reliability or for any other
reason lies exclusively with the authors. This document is not intended as investment
research or investment advice, or a recommendation, offer or solicitation for the
purchase or sale of any security, financial instrument, financial product or service, or
to be used in any way for evaluating the merits of participating in any transaction. 

\small
\begin{spacing}{0.95}
\bibliographystyle{unsrt}
\bibliography{references}

\begin{thebibliography}{10}

\bibitem{lecun2015deep}
Yann LeCun, Yoshua Bengio, and Geoffrey Hinton.
\newblock Deep learning.
\newblock {\em nature}, 521(7553):436--444, 2015.

\bibitem{huang2022partial}
Shudong Huang, Wentao Feng, Chenwei Tang, and Jiancheng Lv.
\newblock Partial differential equations meet deep neural networks: A survey.
\newblock {\em arXiv preprint arXiv:2211.05567}, 2022.

\bibitem{Raissi}
Maziar Raissi.
\newblock Forward-backward stochastic neural networks: Deep learning of
  high-dimensional partial differential equations.
\newblock {\em arXiv preprint arXiv:1804.07010v1}, 2018.

\bibitem{Beck_2019}
Beck Christian, E~Weinan, and Jentzen Arnulf.
\newblock Machine learning approximation algorithms for high-dimensional fully
  nonlinear partial differential equations and second-order backward stochastic
  differential equations.
\newblock {\em Journal of Nonlinear Science}, 29(4):1563--1619, jan 2019.

\bibitem{Han_2018}
Han Jiequn, Jentzen Arnulf, and E~Weinan.
\newblock Solving high-dimensional partial differential equations using deep
  learning.
\newblock {\em Proceedings of the National Academy of Sciences},
  115(34):8505--8510, aug 2018.

\bibitem{E_2017}
E~Weinan, Han Jiequn, and Jentzen Arnulf.
\newblock Deep learning-based numerical methods for high-dimensional parabolic
  partial differential equations and backward stochastic differential
  equations.
\newblock {\em Communications in Mathematics and Statistics}, 5(4):349--380,
  nov 2017.

\bibitem{patel2022quantuminspired}
Raj~G. Patel, Chia-Wei Hsing, Serkan Sahin, Samuel Palmer, Saeed~S. Jahromi,
  Shivam Sharma, Tomas Dominguez, Kris Tziritas, Christophe Michel, Vincent
  Porte, Mustafa Abid, Stephane Aubert, Pierre Castellani, Samuel Mugel, and
  Roman Orus.
\newblock Quantum-inspired tensor neural networks for option pricing.
\newblock {\em arXiv preprint arXiv:2212.14076}, 2022.

\bibitem{Cheridito}
Patrick Cheridito, H~M Soner, Touzi Nizar, and Nicolas Victoir.
\newblock Second-order backward stochastic differential equations and fully
  nonlinear parabolic \uppercase{PDE}s.
\newblock {\em Communications on Pure and Applied Mathematics},
  60(7):1081--1110, nov 2006.

\bibitem{jaimungal}
{\'A}lvaro Cartea, Sebastian Jaimungal, and Jos{\'e} Penalva.
\newblock {\em Algorithmic and high-frequency trading}.
\newblock Cambridge University Press, 2015.

\bibitem{ABOUSSALAH2020112891}
Amine~Mohamed Aboussalah and Chi-Guhn Lee.
\newblock Continuous control with stacked deep dynamic recurrent reinforcement
  learning for portfolio optimization.
\newblock {\em Expert Systems with Applications}, 140:112891, 2020.

\bibitem{hull}
{John C.} Hull.
\newblock {\em Options, Futures, and Other Derivatives, 10th edition}.
\newblock Pearson Prentice Hall, 2017.

\bibitem{wilmott}
Paul Wilmott.
\newblock {\em Paul Wilmott Introduces Quantitative Finance, 2nd Edition}.
\newblock Wiley-Interscience, 2013.

\bibitem{mcdonald2013derivatives}
R.L. McDonald.
\newblock {\em Derivatives Markets: Pearson New International Edition}.
\newblock Pearson Education, 2013.

\bibitem{cheyette2001markov}
Oren Cheyette.
\newblock Markov representation of the heath-jarrow-morton model.
\newblock {\em Available at SSRN 6073}, 2001.

\bibitem{l-s-algo}
Francis Longstaff and Eduardo Schwartz.
\newblock Valuing american options by simulation: A simple least-squares
  approach.
\newblock {\em Review of Financial Studies}, 14:113--47, 02 2001.

\bibitem{Glas04}
Paul Glasserman.
\newblock {\em {Monte Carlo} Methods in Financial Engineering}.
\newblock Springer, New York, NY, USA, 2004.

\bibitem{Raissi-part1}
Maziar Raissi, Paris Perdikaris, and George~Em Karniadakis.
\newblock Physics informed deep learning (part i): Data-driven solutions of
  nonlinear partial differential equations.
\newblock {\em arXiv preprint arXiv:1711.10561}, 2017.

\bibitem{Raissi-part2}
Maziar Raissi, Paris Perdikaris, and George~Em Karniadakis.
\newblock Physics informed deep learning (part ii): Data-driven discovery of
  nonlinear partial differential equations.
\newblock {\em arXiv preprint arXiv:1711.10566}, 2017.

\bibitem{Patel_2022}
Raj~G. Patel.
\newblock Efficient deep learning methods for solving high-dimensional partial
  differential equations for applications in option pricing.
\newblock {\em TSpace Repository, University of Toronto}, Nov 2022.

\bibitem{TNN_NIPS}
Alexander Novikov, Dmitrii Podoprikhin, Anton Osokin, and Dmitry~P Vetrov.
\newblock Tensorizing neural networks.
\newblock {\em Advances in Neural Information Processing Systems}, 28, 2015.

\bibitem{tn_memory}
Karen Simonyan and Andrew Zisserman.
\newblock Very deep convolutional networks for large-scale image recognition.
\newblock {\em International Conference on Learning Representations ({ICLR}),
  2015}, 2015.

\bibitem{xue13_interspeech}
Jian Xue, Jinyu Li, and Yifan Gong.
\newblock {Restructuring of deep neural network acoustic models with singular
  value decomposition}.
\newblock {\em Interspeech, 2013}, pages 2365--2369, 2013.

\bibitem{garciaripoll}
Paula García-Molina, Luca Tagliacozzo, and Juan~José García-Ripoll.
\newblock Global optimization of mps in quantum-inspired numerical analysis.
\newblock {\em arXiv preprint arXiv:2303.09430}, 2023.

\bibitem{li2023quantum}
Yongming Li and Ariel Neufeld.
\newblock Quantum monte carlo algorithm for solving black-scholes pdes for
  high-dimensional option pricing in finance and its proof of overcoming the
  curse of dimensionality.
\newblock {\em arXiv preprint arXiv:2301.09241}, 2023.

\bibitem{jin2022quantum}
Shi Jin, Nana Liu, and Yue Yu.
\newblock Quantum simulation of partial differential equations via
  schrodingerisation: technical details.
\newblock {\em arXiv preprint arXiv:2212.14703}, 2022.

\bibitem{RomanTN}
Roman Orús.
\newblock A practical introduction to tensor networks: Matrix product states
  and projected entangled pair states.
\newblock {\em Annals of Physics}, 349:117--158, 2014.

\bibitem{sukhi_1}
Sukhwinder Singh, Robert N.~C. Pfeifer, and Guifr\'e Vidal.
\newblock Tensor network decompositions in the presence of a global symmetry.
\newblock {\em Phys. Rev. A}, 82:050301, Nov 2010.

\bibitem{tnn_def}
Jacob Biamonte and Ville Bergholm.
\newblock Tensor networks in a nutshell.
\newblock {\em arXiv preprint arXiv:1708.00006}, 2017.

\bibitem{CACIB_1}
Raj Patel, Chia-Wei Hsing, Serkan Sahin, Saeed~S. Jahromi, Samuel Palmer,
  Shivam Sharma, Christophe Michel, Vincent Porte, Mustafa Abid, Stephane
  Aubert, Pierre Castellani, Chi-Guhn Lee, Samuel Mugel, and Roman Orús.
\newblock Quantum-inspired tensor neural networks for partial differential
  equations.
\newblock {\em arXiv preprint arXiv:2208.02235}, 2022.

\bibitem{TNN_Survey}
Maolin Wang, Yu~Pan, Xiangli Yang, Guangxi Li, and Zenglin Xu.
\newblock Tensor networks meet neural networks: A survey.
\newblock {\em arXiv preprint arXiv:2302.09019}, 2023.

\bibitem{NIPS2016_6211}
Stoudenmire Edwin and David~J. Schwab.
\newblock Supervised learning with tensor networks.
\newblock {\em Advances in Neural Information Processing Systems 29}, pages
  4799--4807, 2016.

\bibitem{Stoudenmire_2018}
Stoudenmire Edwin.
\newblock Learning relevant features of data with multi-scale tensor networks.
\newblock {\em Quantum Science and Technology}, 3(3):034003, 2018.

\bibitem{glasser2018supervised}
Ivan Glasser, Nicola Pancotti, and J~Ignacio Cirac.
\newblock Supervised learning with generalized tensor networks.
\newblock {\em arXiv preprint arXiv:1806.05964}, 2018.

\bibitem{efthymiou2019tensornetwork}
Stavros Efthymiou, Jack Hidary, and Stefan Leichenauer.
\newblock Tensor network for machine learning.
\newblock {\em arXiv preprint arXiv:1906.06329}, 2019.

\bibitem{bhatia2019matrix}
Amandeep Bhatia, Mandeep~Kaur Saggi, Ajay Kumar, and Sushma Jain.
\newblock Matrix product state--based quantum classifier.
\newblock {\em Neural computation}, 31(7):1499--1517, 2019.

\bibitem{Liu_2019}
Ding Liu, Shi-Ju Ran, Peter Wittek, Cheng Peng, Raul~Bl{\'{a}}zquez
  Garc{\'{\i}}a, Gang Su, and Maciej Lewenstein.
\newblock Machine learning by unitary tensor network of hierarchical tree
  structure.
\newblock {\em New Journal of Physics}, 21(7):073059, 2019.

\bibitem{9058650}
I.~Glasser, N.~Pancotti, and J.~I. Cirac.
\newblock From probabilistic graphical models to generalized tensor networks
  for supervised learning.
\newblock {\em IEEE Access}, 8:68169--68182, 2020.

\bibitem{PhysRevX.8.031012}
Zhao-Yu Han, Jun Wang, Heng Fan, Lei Wang, and Pan Zhang.
\newblock Unsupervised generative modeling using matrix product states.
\newblock {\em Physical Review X}, 8:031012, 2018.

\bibitem{PhysRevB.99.155131}
Song Cheng, Lei Wang, Tao Xiang, and Pan Zhang.
\newblock Tree tensor networks for generative modeling.
\newblock {\em Physical Review B}, 99:155131, 2019.

\bibitem{PhysRevB.101.075135}
Zheng-Zhi Sun, Cheng Peng, Ding Liu, Shi-Ju Ran, and Gang Su.
\newblock Generative tensor network classification model for supervised machine
  learning.
\newblock {\em Physical Review B}, 101:075135, 2020.

\bibitem{bradley2020modeling}
Tai-Danae Bradley, Edwin~Miles Stoudenmire, and John Terilla.
\newblock Modeling sequences with quantum states: A look under the hood.
\newblock {\em Machine Learning: Science and Technology}, 2020.

\bibitem{quantum_chemistry}
Sandeep Sharma and Ali Alavi.
\newblock Multireference linearized coupled cluster theory for strongly
  correlated systems using matrix product states.
\newblock {\em The Journal of Chemical Physics}, 143(10):102815, sep 2015.

\bibitem{dimensionality}
Ali Zare, Alp Ozdemir, Mark~A. Iwen, and Selin Aviyente.
\newblock Extension of pca to higher order data structures: An introduction to
  tensors, tensor decompositions, and tensor pca.
\newblock {\em arXiv preprint arXiv:1803.00704}, 2018.

\bibitem{subspace}
Jing Zhang, Xinhui Li, Peiguang Jing, Jing Liu, and Yuting Su.
\newblock Low-rank regularized heterogeneous tensor decomposition for subspace
  clustering.
\newblock {\em IEEE Signal Processing Letters}, 25(3):333--337, 2018.

\end{thebibliography}
\end{spacing}

\end{document}